\documentclass[12pt,preprint]{aastex}

\def\beq{\begin{eqnarray}}
\def\eeq{\end{eqnarray}}

\def\Qvis{Q_{\rm vis}}
\def\Qadv{Q_{\rm adv}}
\def\Qphotodis{Q_{\rm photodis}}
\def\Qnu{Q_{\rm \nu}}
\def\Msuns{$M_\sun$ s$^{-1}$}

\def\Ye{$Y_{\rm e}$}
\def\Yp{$Y_{\rm p}$}
\def\Yn{$Y_{\rm n}$}
\def\Yalpha{$Y_{\rm {\alpha}}$}
\def\Xnuc{$X_{\rm {nuc}}$}

\begin{document}

\title{Structure and Luminosity of Neutrino-cooled Accretion Disks}

\author{Tong Liu, Wei-Min Gu, Li Xue, and Ju-Fu Lu*}

\affil{Department of Physics
and Institute of Theoretical Physics and Astrophysics, \\
Xiamen University, Xiamen, Fujian 361005, China}

\email{*lujf@xmu.edu.cn}

\begin{abstract}
 Neutrino-cooled hyperaccretion disks around stellar mass black holes are plausible
 candidates for the central engine of gamma-ray bursts. We calculate the one-dimensional
 structure and the annihilation luminosity of such disks. The neutrino optical depth is
 of crucial importance in determining the neutrino cooling rate and is in turn dependent on the
 electron fraction, the free nucleon fraction, and the electron degeneracy, with
 given density and temperature of the disk matter. We construct a bridging formula
 for the electron fraction that works for various neutrino optical depths, and give
 exact definitions for the free proton fraction and free neutron fraction. We show
 that the electron degeneracy has important effects in the sense that it enlarges
 the absorption optical depth for neutrinos, and it along with the neutronization
 processes favored by high temperature cause the electron fraction to drop to be
 below 0.1 in the inner region of the disk. The resulting neutrino annihilation
 luminosity is considerably reduced comparing with that obtained in previous works
 where the electron degeneracy was not considered and the electron fraction was simply
 taken to be 0.5, but it is still likely to be adequate for gamma-ray bursts, and it
 is ejected mainly from the inner region of the disk and has an anisotropic distribution.
\end{abstract}


\section{Introduction}

Theoretical models for gamma-ray bursts (GRBs) can be divided into
two categories: those named fireball models that treat the shock
interaction of relativistic outflows and production of gamma rays
and afterglows in other wavelengths (see, e.g., M\'esz\'aros 2002
and Zhang \& M\'esz\'aros 2004 for reviews), and those that explore
the central engine of the fireball, i.e., the energy source of
relativistic outflows. Most popular models in the latter category
are in common invoking a stellar-mass black hole accreting with a
hypercritical rate, of the order of 1 {\Msuns}. The main problem in
these models is how to convert some fraction of the released
gravitational energy of the accreted matter into a relativistic
outflow, creating an explosion with energy $\sim$
$10^{50}$-$10^{52}$ ergs (depending on whether emission is isotropic
or it is beamed). Two mechanisms have been proposed to tackle this
problem: the neutrino emission and annihilation, and the energy
extraction from the accretion disk and/or the black hole via
magnetohydrodynamical processes (see, e.g., Popham et al. 1999 and
Di Matteo et al. 2002 for references). The former mechanism is
easier to understand and can be calculated more accurately, and is
the topic we wish to discuss in this paper.

    In the inner region of such a hyperaccretion disk the density and
temperature are so high ($\rho \sim$ $10^{10}$ g~cm$^{-3}$, $T \sim
10^{10}$ K) that photons are totally trapped, and large amounts of
energetic neutrinos are emitted, carrying away the viscous
dissipation energy of accreted gas. Annihilation of some fraction of
emitted neutrinos produces a relativistic electron-positron outflow
to power a GRB. The properties of such a neutrino-cooled accretion
disk model were first worked out in details by Popham et al. (1999).
From the observational point of view, the key question to be
answered in this model is whether the neutrino annihilation can
indeed provide sufficient energy for GRBs. Popham et al. (1999) gave
a positive answer to this question, but they assumed a priori that
the accretion disk is transparent for neutrinos, thus their neutrino
radiation luminosity (before annihilation) $L_{\nu}$ and accordingly
neutrino annihilation luminosity $L_{\nu \overline{\nu}}$ might be
overestimated. Later, Di Matteo et al. (2002) recalculated $L_{\nu}$
and $L_{\nu \overline{\nu}}$ in the neutrino-cooled accretion disk
model, taking the neutrino opacity into account. They obtained that
even for a modest mass accretion rate the inner region of the disk
becomes opaque and neutrinos are sufficiently trapped, the resulting
$L_{\nu \overline{\nu}}$ is $\lesssim$ $10^{50}$ ergs s$^{-1}$ and
is inadequate for GRBs. In a recent work (Gu et al. 2006) we showed
that when the general relativistic effect is considered and the
contribution from the optically thick region is included, the
neutrino-cooled accretion disk can work as the central engine for
GRBs, although the correct $L_{\nu \overline{\nu}}$ is somewhat
lower than that of Popham et al. (1999) because of the effect of
neutrino opacity.

   In addition to the general relativity and the neutrino opacity,
there are certainly other factors that may influence the neutrino
radiation and annihilation of a neutrino-cooled accretion disk, such
as the electron degeneracy and the electron fraction. In our
previous work (Gu et al. 2006) the treatment of electron degeneracy
was oversimplified, and the electron fraction $Y_{\rm e}$ was simply
taken to be 0.5, i.e., assuming an equal mix of protons and neutrons
since ${Y_{\rm e}}$ = ${n_{\rm p}}$/(${n_{\rm p}}$ + ${n_{\rm n}}$),
where ${n_{\rm p}}$ and ${n_{\rm n}}$ are the total number density
of protons and that of neutrons, respectively. Kohri \& Mineshige
(2002) pointed out that when electrons are degenerate, there emerges
an important consequence that the electron-positron pair creation
and accordingly the neutrino emission are suppressed. Kohri et al.
(2005) took great care to calculate the ratio of free neutrons to
free protons $\tilde{n}_{\rm n}$$/$$\tilde{n}_{\rm p}$ since this
ratio has a large effect on the neutrino emission rates: the true
ratio $\tilde{n}_{\rm n}$$/$$\tilde{n}_{\rm p}$$>$ 1 (or $Y_{\rm e}
< 0.5$) resulted from neutronization processes will also lead to a
suppression of neutrino emission (note that they did not distinguish
between ${n_{\rm n}}$, ${n_{\rm p}}$ and $\tilde{n}_{\rm
n}$,$\tilde{n}_{\rm p}$, see \S 2.3.1). They made calculations for
the neutrino emission rates and other quantities even in the
delicate regime where the electron degeneracy is moderate, which is
also a significant improvement over previous works. Lee et al.
(2005) also considered the effects of electron degeneracy and
electron fraction in neutrino-cooled accretion disks: they used an
expression for the pressure of ultra-relativistic electrons with
arbitrary degeneracy and calculated ${Y_{\rm e}}$ with an
approximate bridging formula that allows for a transition from the
neutrino optically thin to optically thick regime. These works
(Kohri \& Mineshige 2002; Kohri et al. 2005; Lee et al. 2005) made
advances in microphysics, but they were within the Newtonian
framework. Very recently, Chen \& Beloborodov (2006) presented fully
relativistic calculations of the structure of neutrino-cooled
accretion disks around Kerr black holes and proved that both the
electron degeneracy and the electron fraction dramatically affect
the disk and its neutrino emission.

   It is seen from the above brief review that the electron degeneracy
and the lower electron fraction certainly suppress the neutrino
emission and reduce the neutrino annihilation luminosity of a
neutrino-cooled accretion disk. The question remains whether the
reduced neutrino annihilation luminosity is still adequate for GRBs.
In this paper we try to refine our previous results of the structure
and luminosity of neutrino-cooled accretion disks by considering the
relevant microphysics more completely and more accurately.

\section{Physics of Neutrino-cooled Accretion Disks}

\subsection{Hydrodynamics}

   We limit the central accreting black hole to be a non-rotating one,
its general relativistic effect is simulated by the well-known
Paczy\'nski \& Wiita (1980) potential $\Phi=-GM/(R-R_{g})$, where
$M$ is the black hole mass, $R$ is the radius, and $R_{g}=2GM/c^ 2$
is the Schwarzschild radius.

   As in the relevant previous works, the hydrodynamics of
hyperaccretion disks is expected to be similar to that of normal
accretion disks in X-ray binaries, and is well approximated by that
of steady axisymmetric height-averaged accretion flows (e.g., Chap.
3 of Kato et al. 1998). Accretion in the disk is driven by viscous
stress, and the kinematic viscosity coefficient is expressed as
$\nu={\alpha}c_{\rm s} H$, where $H$ is the half thickness of the
disk; $c_{\rm s}=(P/\rho)^{1/2}$ is the isothermal sound speed, with
$P$ and $\rho$ being the pressure and mass density, respectively;
and $\alpha$ is a dimensionless constant parameter that absorbs all
the detailed microphysics of viscous processes. The angular velocity
is approximately Keplerian, i.e., $\Omega=\Omega_{\rm K}=(GM /R)^{1/
2}/(R - R_{g})$. The disk is in the vertical hydrostatic
equilibrium, and this gives $H=c_{\rm s}/\Omega_{\rm K}$. With these
simplifications the problem is reduced to be one-dimensional, i.e.,
all physical quantities depend on $R$ only.

   The constant mass accretion rate $\dot{M}$ is expressed from the
continuity equation as \beq \ \dot{M}=-4{\pi}RH {\rho} v,\ \eeq
where $v$ is the radial velocity that can be read from the angular
momentum equation as \beq\ v=\alpha H c_{\rm s}(1 - j/ \Omega_{\rm
K} R^ 2)^{- 1} \frac{d\ln \Omega_{\rm K}}{d R} ,\ \eeq where $j$ is
an integration constant determined by the zero-torque boundary
condition at the last stable orbit, and it represents the specific
angular momentum (per unit mass) of the matter accreted into the
black hole.

\subsection{Thermodynamics}

   The energy equation is generally written as the balance between the
viscous heating and the cooling rates (per unit area of a half-disk
above or below the equator), \beq\ \Qvis=Q^{-}.\ \eeq The viscous
heating rate $\Qvis$ is similar to that of normal accretion disks,
\beq\ \ \Qvis=- \frac{1}{4\pi} \dot{M} \Omega_{\rm K} R
(1-j/\Omega_{\rm K} R^2) \frac{d\Omega_{\rm K}}{d R} .\ \eeq But the
cooling rate $Q^{-}$ is crucially different, it has three
contributions: \beq\ \ Q^{-}=\Qphotodis + \Qadv +\Qnu .\ \eeq In
this equation there is no cooling term of photon radiation (it is
practically zero in our calculations). Instead, photons are totally
trapped in the disk, so they contribute to the advective cooling and
the pressure (see below). The cooling rate by photodisintegration of
$\alpha$-particles $\Qphotodis$ is \beq\ \ \Qphotodis = 6.8 \times
10^{28} \rho_{10} v H \frac{d X_{\rm nuc}}{d R}~{\mathrm
{cgs~units}},\ \eeq where ${\rho}_{10} \equiv {\rho}/10^{10}{\rm
g~cm^{-3}}$, and $X_{\rm nuc}$ is the mass fraction of free nucleons
(e.g., Kohri et al. 2005). The advective cooling rate $\Qadv$ is
\beq\ \ \Qadv = \rho v H T \frac{d s}{d R} \approx - \xi v
\frac{H}{R} T ( \frac{4}{3} a T^3 +\frac{3}{2} \frac{k_B \rho}{m_u}
\frac{1+3X_{\rm nuc}}{4} + \frac{4}{3} \frac{u_{\rm \nu}}{T}),\ \eeq
where $s$ is the specific entropy, $ \xi\propto- d \ln s/d \ln R $
is taken to be equal to 1, i.e., $d s/d R$ is approximated as $s /
R$ (Kohri \& Mineshige 2002), and $m_u$ is the mean mass of a
nucleon. The entropy of degenerate particles is small and can be
neglected. The three terms in the brackets of equation (7) are the
entropy density of photons, of free nucleons and $\alpha$-particles,
and of neutrinos, respectively; and ${u}_{\nu}$ is the energy
density of neutrinos, for which we adopt a bridging formula valid in
both the optically thin and thick regimes (Popham \& Narayan 1995;
Di Matteo et al. 2002),\beq\  u_{\rm \nu}=\sum_{i} \frac{(7/8)a T^4
(\tau_{{\nu}_i}/2+1/ \sqrt{3})} {\tau_{{\nu}_i}/2+1/ \sqrt{3}+1/(3
 \tau_{a,{\nu}_i})},  \eeq where $\tau_{{\nu}_i}$ is the total
optical depth for neutrinos, $\tau_{a,{\nu}_i}$ is the absorption
optical depth for neutrinos, and the subscript $i$ runs for the
three species of neutrinos $\nu_{\rm e}$ , $\nu_{\rm \mu}$ , and
$\nu_{\rm \tau}$ . The cooling rate due to neutrino loss $Q_{\rm
\nu}$ is expressed in accordance with the above equation(Kohri et
al. 2005),\beq\ Q_{\nu}=\sum_{i} \frac{(7/8) {\sigma} T^4 }
{(3/4)[\tau_{{\nu}_i}/2+1/ \sqrt{3}+1/(3
 \tau_{a,{\nu}_i})]}.  \eeq

   The equation of state is also very different from that of normal
accretion disks, as the contributions to the pressure from
degenerate electrons (we assume that nucleons are not degenerate
throughout the present paper) and from neutrinos should be included,
it is written as \beq\  P=P_{\rm gas}+P_{\rm rad}+P_{\rm e}+P_{\rm
\nu}.  \eeq The gas pressure from free nucleons and
$\alpha$-particles $P_{\rm gas}$ is \beq\  P_{\rm gas}=\frac{k_{\rm
B} \rho T}{m_u} \frac{1+3 X_{\rm nuc}}{4} . \eeq The photon
radiation pressure $P_{\rm rad}$ is \beq\ P_{\rm rad}=aT^4/3. \eeq
The electron pressure $P_{\rm e}$ is from both electrons and
positrons and should be calculated using the exact Fermi-Dirac
distribution. No asymptotic expansions are valid because at
different radii electrons may be with different degrees of
degeneracy and may be relativistic or nonrelativistic. It reads
\beq\ P_{\rm e}=P_{\rm e^{-}}+ P_{\rm e^{+}}, \eeq with \beq\ P_{\rm
e^{\mp}}= \frac{1}{3 {\pi}^2 {\hbar}^3 c^3} \int_0^{\infty} d p
\frac{p^4}{\sqrt{p^2 c^2+{m_{\rm e}}^2 c^4}} \frac{1}{e^{({\sqrt{p^2
c^2+{m_{\rm e}}^2 c^4} \mp {\mu_{\rm e}})/k_{\rm B} T}}+1}, \eeq
where $\mu_{\rm e}$ is the chemical potential of electrons, and the
electron degeneracy is measured by the degeneracy parameter defined
as $\eta_{\rm e} = \mu_{\rm e} / k_{\rm B} T$. We agree with Lee et
al. (2005) that since the presence of relativistic $e^{-} e^{+}$
pairs is automatically taken into account in the expression for
$P_{\rm e}$, there is no alteration to the numerical factor 1/3 in
the expression for $P_{\rm rad}$, nor to the factor 4/3 for the
photon entropy in equation (7). The neutrino pressure $P_{\rm \nu}$
is \beq\ P_{\rm \nu}=u_{\rm \nu} / 3. \eeq

\subsection{Microphysics}

   All the equations in the above two subsections can be combined into
only two equations, i.e., equations (1) and (3). In these two
equations there are seven unknown quantities, namely $\rho$, $T$,
\Xnuc, $\tau_{\nu_i}$($\nu_i = \nu_e, \nu_\mu,\nu_\tau$), and
$\mu_{\rm e}$(or $\eta_{\rm e}$). Therefore, one has to find more
equations relating these unknowns with the knowledge of
microphysics.

\subsubsection{Neutrino optical depth}

The total optical depth for neutrinos is \beq\ \tau_{\nu_i}=\tau_{s,
\nu_i} + \tau_{a,\nu_i}. \eeq

The optical depth for neutrinos through scattering off free
nucleons, $\alpha$-particles, and electrons $\tau_{s, \nu_i}$ is
given by \beq\ \tau_{s, \nu_i}=H/\lambda_{\nu_i}=H [\sigma_{p,\nu_i}
\tilde{n}_{\rm p}+\sigma_{n,\nu_i} \tilde{n}_{\rm
n}+\sigma_{\alpha,\nu_i} n_{\rm \alpha}+\sigma_{e,\nu_i} (n_{\rm
e^{-}}+n_{\rm e^{+}})], \eeq where $\lambda_{\nu_i}$ is the mean
free path; $\sigma_{p,\nu_i}$ , $\sigma_{n,\nu_i}$ ,
$\sigma_{\alpha,\nu_i}$ and $\sigma_{e,\nu_i}$
 are the cross sections of scattering on protons, neutrons,
$\alpha$-particles, and electrons; $\tilde{n}_{\rm p}$ ,
$\tilde{n}_{\rm n}$ , $n_{\rm \alpha}$ , $n_{\rm e^{-}}$ and $n_{\rm
e^{+}}$ are the number densities of free protons, free neutrons,
$\alpha$-particles, electrons, and positrons, respectively. The four
cross sections are given by (Burrows \& Thompson 2002) \beq\
\sigma_{p,\nu_i}=\frac{\sigma_0 E_{\nu_i}^2}{4}[(C_{V,\nu_i}-1)^2+3
g_A^2 (C_{A,\nu_i}-1)^2] ,\eeq\ \beq\
\sigma_{n,\nu_i}=\frac{\sigma_0 E_{\nu_i}^2}{4} \frac{1+3 g_A^2}{4},
\eeq\ \beq\ \sigma_{\alpha,\nu_i}=4 \sigma_0 E_{\nu_i}^2 \sin^4
\theta_W ,\eeq\  \beq\ \sigma_{e,\nu_i}= \frac{3 k_{\rm B} T
\sigma_0 E_{\nu_i}}{8 m_{\rm e} c^2} (1+ \frac{\eta_{\rm
e}}{4})[(C_{V,\nu_i}+C_{A,\nu_i})^2+ \frac{1}{3}
(C_{V,\nu_i}-C_{A,\nu_i})^2],\eeq where $\sigma_0= 1.76 \times 10^{-
44} {\rm cm^2}$,$E_{\nu_i}$ is the mean energy of neutrinos in units
of $(m_{\rm e} c^2)$, $g_A \approx 1.26$, $\sin^2\theta_W \approx
0.23$, $C_{V,\nu_e}=1/2 + 2 \sin^2 \theta_W$ , $C_{V,\nu_\mu}
=C_{V,\nu_\tau}= - 1/2 + 2sin^2 \theta_W$ , $ C_{A,\nu_e}=
C_{A,\overline{\nu}_\mu}=C_{A,\overline{\nu}_\tau}=1/2$ ,
$C_{A,\overline{\nu}_e} = C_{A,\nu_\mu}=C_{A,\nu_\tau}=- 1/2$. Since
$\tilde{n}_{\rm p}={n}_{\rm p}-2 n_{\rm \alpha}$ and $\tilde{n}_{\rm
n} = n_{\rm n} - 2 n_{\rm \alpha}$, the free proton fraction $Y_{\rm
p}$, the free neutron fraction \Yn, and the $\alpha$-particle
fraction $Y_{\rm \alpha}$ are related to $Y_{\rm e}$  and $X_{\rm
nuc}$ as \beq\
 Y_{\rm p}= \frac{\tilde{n}_{\rm p}}{n_{\rm p}+n_{\rm n}}=Y_{\rm e}-
\frac{1-X_{\rm nuc}}{2}, \eeq\ \beq\ Y_{\rm n}=\frac{\tilde{n}_{\rm
n}}{{n}_{\rm p}+{n}_{\rm n}}=1-Y_{\rm e}- \frac{1-X_{\rm nuc}}{2},
\eeq\ \beq\ Y_{\rm \alpha}= \frac{n_{\rm \alpha}}{n_{\rm p}+n_{\rm
n}}= \frac {1- X_{\rm nuc}}{4}, \eeq  and $n_{\rm e^{-}}$, $n_{\rm
e^{+}}$ are given by the Fermi-Dirac integration, \beq\ n_{\rm
e^{\mp}}= \frac{1}{\hbar^3 \pi^2} \int_0^\infty d p~p^2
\frac{1}{e^{({\sqrt{p^2 c^2+{m_{\rm e}}^2 c^4} \mp {\mu_{\rm
e}})/k_{\rm B} T}}+1} \eeq (cf. Kohri et al. 2005, they took
$\tilde{n}_{\rm p}= {n}_{\rm p}$ and $\tilde{n}_{\rm n} ={n}_{\rm
n}$, and thus $Y_{\rm p} = Y_{\rm e}$ and $Y_{\rm n} = 1 - Y_{\rm
e}$, which are valid only for the completely dissociated matter,
i.e., $X_{\rm nuc} = 1$).

   The absorption depth for neutrinos $\tau_{a, \nu_i}$ is defined
by \beq\ \tau_{a, \nu_i}= \frac{q_{\nu_i} H}{4 (7/8) \sigma
T^4},\eeq where $q_{\nu_i}$ is the total neutrino cooling rate (per
unit volume) and is the sum of four terms, \beq\ q_{\nu_i}=q_{\rm
 URCA}+q_{\rm e^{-} e^{+}}+q_{\rm brem}+q_{\rm plasmon}.\eeq The
neutrino cooling rate due to the URCA processes $q_{\rm URCA}$
relates only to $\nu_{\rm e}$ and is represented by the sum of three
terms (Chap. 11 of Shapiro \& Teukolsky 1983; Yuan 2005), \beq\
q_{\rm URCA}=q_{p+e^{-}\rightarrow n+\nu_{\rm
e}}+q_{n+e^{+}\rightarrow p+\overline{\nu}_{\rm e}}+q_{n \rightarrow
p+e^{-}+\overline{\nu}_{\rm e}}, \eeq with \beq\
q_{p+e^{-}\rightarrow n+\nu_{\rm e}} = \frac{G_F^2 \cos^2
\theta_c}{2 \pi^2 \hbar^3 c^2}(1+3 g_A^2) \tilde{n}_{\rm p}
\int_Q^\infty d E_e~E_e~\sqrt{{E_e}^2-{m_{\rm e}}^2 c^4}(E_e-Q)^3
f_{e^{-}} ,\eeq \beq\ q_{n+e^{+}\rightarrow p+\overline{\nu}_{\rm
e}}=\frac{G_F^2 \cos^2 \theta_c}{2 \pi^2 \hbar^3 c^2}(1+3 g_A^2)
\tilde{n}_{\rm n} \int_{m_{\rm e} c^2}^\infty d
E_e~E_e~\sqrt{{E_e}^2-{m_{\rm e}}^2 c^4}(E_e+Q)^3 f_{e^{+}}, \eeq
\beq\ q_{n \rightarrow p+e^{-}+\overline{\nu}_{\rm e}}=\frac{G_F^2
\cos^2 \theta_c}{2 \pi^2 \hbar^3 c^2}(1+3 g_A^2) \tilde{n}_{\rm n}
\int_{m_{\rm e} c^2}^Q d E_e~E_e~\sqrt{{E_e}^2-{m_{\rm e}}^2
c^4}(Q-E_e)^3 (1-f_{e^{-}}), \eeq where $G_F \approx 1.436 \times
10^{-49} {\rm ergs~cm^3}$, $\cos^2 \theta_c \approx 0.947$, $Q =
(m_{\rm n} - m_{\rm p})c^2$, and $f_{e^{\mp}}=\{\exp[(E_{\rm e} \mp
\mu_{\rm e})/k_{\rm B} T]+1\}^{-1}$is the Fermi-Dirac function. The
last term in the right hand side of equation (28), i.e., equation
(31), is small comparing with the other two terms, and was usually
not included in the literature. The electron-positron pair
annihilation rate into neutrinos $q_{\rm e^{-}e^{+}}$  is (e.g.,
Itoh et al. 1989) \beq\ q_{e^{-}e^{+} \rightarrow \nu_{\rm e}
\overline {\nu}_{\rm e}} \approx 3.4 \times 10^{33} T_{11}^9 ~{\rm
ergs~cm^{-3}~s^{-1}},\eeq \beq\ q_{e^{-}e^{+} \rightarrow \nu_{\rm
\mu} \overline {\nu}_{\rm \mu}}=q_{e^{-}e^{+} \rightarrow \nu_{\rm
\tau} \overline {\nu}_{\rm \tau}} \approx 0.7 \times 10^{33}
T_{11}^9 ~{\rm ergs~cm^{-3}~s^{-1}},\eeq where $T_{11} \equiv
T/10^{11} {\rm K}$. Expressions (32) and (33) are valid in the
nondegenerate limit $\eta_{\rm e} \ll 1$, and when electrons are
degenerate $q_{\rm e^{-} e^{+}}$ becomes negligible. The
nucleon-nucleon bremsstrahlung rate $q_{\rm brem}$ through the
processes $n + n \rightarrow  n + n + \nu_i+\overline{\nu}_i$ is the
same for the three species of neutrinos (Hannestad \& Raffelt 1998;
Burrows et al. 2000),\beq\ q_{\rm brem} \approx 1.5 \times 10^{27}
\rho_{10}^2 T_{11}^{5.5} ~{\rm ergs~cm^{-3}~s^{-1}}.\eeq As to the
plasmon decay rate $q_{\rm plasmon}$, only that through the process
$\tilde{\gamma} \rightarrow \nu_{\rm e}+\overline \nu_{\rm e}$ needs
to be considered, where plasmons $\tilde{\gamma}$ are photons
interacting with electrons, \beq\ q_{\rm plasmon} \approx 1.5 \times
10^{32} T_{11}^9 \gamma_p^6 \exp {(-\gamma_p)} (1+ \gamma_p) (2+
\frac{\gamma_p^2}{1+\gamma_p})~{\rm ergs~cm^{-3}~s^{-1}},\eeq where
$\gamma_p= 5.565 \times 10^{-2} {[(\pi^2 + 3
\eta_e^2)/3]}^{1/2}$(Ruffert et al. 1996). It is expected that
$q_{\rm brem}$ and $q_{\rm plasmon}$ can become important only at
very high electron degeneracy.

\subsubsection{Electron Fraction}

It is seen that the electron fraction $Y_{\rm e}$ mentioned in
Introduction has appeared in the expression of neutrino optical
depth $\tau_{\nu_i}$. In order to calculate $\tau_{\nu_i}$  and
accordingly the neutrino cooling rate $\Qnu$ (see eq. [9]), more
knowledge about $Y_{\rm e}$  is required.

   Beloborodov (2003) proved that $\beta$-equilibrium among free neutrons,
free protons, and electrons is established in disks with $\dot{M}
\gtrsim 10^{31} {(\alpha/0.1)}^{9/5}(M/M_\sun)^{6/5}~ {\rm g
~s^{-1}}$, a condition that is likely to be fulfilled for
hyperaccretion disks; and noted that a distinction needs to be made
to determine the equilibrium composition depending on the optical
depth of the disk material. Kohri et al.(2005) discussed various
timescales in neutrino-cooled accretion disks in details, and showed
that for sufficiently large $\dot{M}$ and not too large $R$ the
timescale for the reactions from proton to neutron and from neutron
to proton is shorter than the dynamical timescale (the accretion
time), i.e., the $\beta$-equilibrium is likely to be realized.

If the disk material is opaque to neutrinos, there are reversible
reactions ${e}^{-}+{p} \rightleftharpoons {n}+\nu_{\rm e}$ and
${e}^{+} + {n}\rightleftharpoons {p}+\overline{\nu}_{\rm e}$. The
chemical potential of neutrinos can be ignored because the number
density of neutrinos and that of antineutrinos are likely to be
equal, then the $\beta$-equilibrium condition is \beq\ \mu_{\rm
n}=\mu_{\rm p}+\mu_{\rm e}, \eeq where $\mu_{\rm n}$ and $\mu_{\rm
p}$ are chemical potentials of neutrons and protons, respectively.
On the other hand, if the material is transparent to neutrinos,
there are no reversible reactions as for the neutrino opaque
material, but the $\beta$-equilibrium can be reached when the rate
of reaction ${e}^{-}+{p} \rightarrow {n}+\nu_{\rm e}$ is equal to
that of reaction ${e}^{+} + {n}\rightarrow {p}+\overline{\nu}_{\rm
e}$. Yuan(2005) calculated these two rates and obtained \beq\
\mu_{\rm n}=\mu_{\rm p}+2 \mu_{\rm e}. \eeq The neutron-to-proton
ratio in $\beta$-equilibrium is given by \beq\ \frac{\tilde {n}_{\rm
n}}{\tilde {n}_{\rm p}}=\exp {[(\mu_{\rm n}-\mu_{\rm p}-Q)/k_{\rm B}
T]},\eeq which results in \beq\ \lg{\frac{\tilde {n}_{\rm n}}{\tilde
{n}_{\rm p}}}=\frac{\mu_{\rm e}-Q}{k_{\rm B} T},\eeq for the
neutrino opaque limit, and \beq\ \lg{\frac{\tilde {n}_{\rm
n}}{\tilde {n}_{\rm p}}}=\frac{2 \mu_{\rm e}-Q}{k_{\rm B} T},\eeq
for the neutrino transparent limit. In order to allow for a
transition from the optically thin to optically thick regime, we
adopt a treatment similar to that in Lee et al. (2005), i.e.
introducing a weight factor $f(\tau_\nu) = \exp (-{\rm
\tau}_{\nu_{\rm e}} )$ and writing in a combined form,\beq\
\lg{\frac{\tilde {n}_{\rm n}}{\tilde {n}_{\rm p}}}= f(\tau_\nu)
\frac{2 \mu_{\rm e}-Q}{k_{\rm B} T}+[1-f(\tau_\nu)] \frac{\mu_{\rm
e}-Q}{k_{\rm B} T}. \eeq By using the relation $Y_{\rm e} = n_{\rm
p}/(n_{\rm p} + n_{\rm n})$ and equations (22), (23), and (24) we
finally arrive at \beq\ Y_{\rm e}=\frac{1}{2}(1-X_{\rm nuc})+
\frac{X_{\rm nuc}}{1+\exp\{\frac{[1+f(\tau_\nu)] \mu_{\rm
e}-Q}{k_{\rm B } T}\}}. \eeq Some comments on equation (42) are in
order. First, this equation is more exact than equation (12) of Lee
et al. (2005), theirs is the first term of the expansion of ours.
Second, we notice that Kohri et al. (2005) discussed the ratio
$\tilde{n}_{\rm n}/\tilde{n}_{\rm p}$ at length. They started with
considering the transition rates from proton to neutron and from
neutron to proton, then obtained the expression of $\tilde{n}_{\rm
n}/\tilde{n}_{\rm p}$ , i.e., equation (39) for the neutrino opaque
limit, and adopted an approximate procedure to estimate this ratio
for the neutrino optically thin regime. We stress that equation (40)
is rigorous for the neutrino transparent limit, since it is also
derived from the transition rates between protons and neutrons (Yuan
2005); then we construct equation (42) as a bridging formula between
the opaque and transparent limits and expect it to work for the
regime where the neutrino optical depth is moderate.

An additional relation between $Y_{\rm e}$   and $X_{\rm nuc}$  can
be found from the equation of nuclear statistical equilibrium (e.g.,
Meyer 1994), \beq\ Y(Z,A)  = G(Z,A)[{\zeta(3)}^{A-1} \pi^{(1-A)/2}
2^{(3A-5)/2}] A^{3/2} {(\frac {k_{\rm B}T}{m_u c^2})}^{3(A-1)/2}
\phi^{1-A} {Y_{\rm p}}^Z {Y_{\rm n}}^{A-Z} \exp{[\frac
{B(Z,A)}{k_{\rm B} T}]}, \eeq where $Y(Z,A)$ is the mass fraction of
a kind of particles with the charge number $Z$ and mass number $A$,
$G(Z, A)$ is the nuclear partition function, $\zeta(3)$ is the
Riemann zeta function of argument 3, $\phi = [2\zeta (3)(k_{\rm B}
T)^3]/ [\pi^2 (\hbar c)^3 \rho N_A]$ is the photon-to-baryon ratio,
$N_A$ is the Avagadro constant, and $B(Z, A)$ is the binding energy
of the nucleus. As in all the previous works, we assume that all
heavy nuclei are $\alpha$-particles. This should be a reasonable
assumption because all nuclei heavier than $\alpha$-particles
contain approximately equal numbers of neutrons and protons. Then by
using equations (22), (23), and (24), equation (43) yields \beq\
4(Y_{\rm e}- \frac {1-X_{\rm nuc}}{2})(1-Y_{\rm e}- \frac {1-X_{\rm
nuc}}{2}){(1-X_{\rm nuc})}^{-1/2}={8.71 \times 10^4 \rho_{10}^{-3/2}
T_{11}^{9/4} \exp{(-1.64/T_{11})}}.\eeq Only Chen \& Beloborodov
(2006) adopted this equation, while other previous works used a
simple expression of $X_{\rm nuc}$ that was obtained by taking
$Y_{\rm e} = 0.5$ and did not reflect the interdependence between
$Y_{\rm e}$  and $X_{\rm nuc}$ (e.g., eq. [40] of Kohri et al.
2005).

\subsubsection{Electron Chemical Potential}

The electron chemical potential is determined by the condition of
charge neutrality among protons, electrons, and positrons, \beq\
n_{\rm p}= \frac {\rho Y_{\rm e}}{m_u} = n_{\rm e^{-}}-n_{\rm
e^{+}},\eeq with $n_{\rm e^{-}}$ and $n_{\rm e^{+}}$  given in
equation (25).

\subsection{Summary of Equations}

The system of equations is closed, as there are eight equations,
i.e., equations (1), (3), (42), (44), (45), and three equations (16)
for eight independent unknowns $\rho$,$T$,$\tau_{\nu_{\rm
e}}$,$\tau_{\nu_{\rm \mu}}$,$\tau_{\nu_{\rm \tau}}$,$X_{\rm nuc}$,
$Y_{\rm e}$,and $\mu_{\rm e}$(or $\eta_{\rm e}$), which can be
numerically solved as functions of $R$ with given constant
parameters $M$,$\dot{M}$,$\alpha$, and $j$. All the other quantities
such as $P$ and its components, $Q^{-}$ and its components, and
composition fractions \Yp , \Yn , and $Y_{\rm \alpha}$   are
obtained accordingly.

Our equation set has the following advantages: (1) Effects of
relevant factors are taken into account in a combined way, including
the general relativity, the inner boundary condition of the disk,
various processes that contribute to the neutrino cooling and the
neutrino opacity, the electron degeneracy, the electron fraction,
and the coexistence of electrons, positrons, free protons, free
neutrons, and $\alpha$-particles. (2) Whenever the electron
degeneracy is concerned we use the exact Fermi-Dirac integration
rather than the analytical approximations that are valid only for
extreme cases. (3) We take great care to calculate the neutrino
optical depth and the electron fraction. In doing so, we make a
careful distinction between the total nucleon number densities
$n_{\rm n}$, $n_{\rm p}$ and the free nucleon number densities
$\tilde {n}_{\rm n}$,$\tilde {n}_{\rm p}$, so that the composition
of the disk matter can be exactly described by fractions $Y_{\rm e}$
, $Y_{\rm p}$ , $Y_{\rm n}$ , and $Y_{\rm \alpha}$ ; we propose a
new bridging formula for $Y_{\rm e}$  (eq. [42]) from the
$\beta$-equilibrium condition, which is applicable to both the
neutrino optically thin and optically thick regimes; and we adopt
equation (44) that connects $Y_{\rm e}$ with \Xnuc.

\section{Numerical Results for The Disk Structure}

This section presents our results for the structure of a
neutrino-cooled accretion disk, obtained by numerically solving the
set of eight equations in \S 2. Figures 1 - 6 show physical
quantities of the disk matter as functions of $R$. In all these
figures the necessary constant parameters are fixed to be their most
typical values, i.e., $M = 3{M_\sun}$, $\dot{M}$= 1 {\Msuns} (except
for Fig. 5 where $\dot{M}$= 5\Msuns),$\alpha = 0.1$, and $j = 1.8 c
R_g$ (see, e.g., Gu et al. 2006 for the discussion of $j$).

Figure 1 is for the density $\rho$, temperature $T$, and electron
degeneracy parameter $\eta_{\rm e}$. It is seen that from $R = 500
R_g$ inward to $R = 3R_g$, $\rho$ increases by about four orders of
magnitudes (Fig. 1a), $T$ increases by about eight times (Fig. 1b);
and in the innermost region of the disk $\rho$ reaches to $\sim
10^{11} ~{\rm g~ cm^{-3}}$, and $T$ reaches to $\sim 3 - 4 \times
10^{10} {\rm K}$. With such densities and temperatures, $\eta_{\rm
e}$ increases first with decreasing $R$, reaches to its maximum
value at $R\sim 65 R_g$, and then decreases with decreasing $R$
because of increasing temperature (Fig. 1c). The behavior of
$\eta_{\rm e}$ obtained by us is consistent with that of Kohri et
al. (2005, Fig. 4 there); but is not so with that of Chen \&
Beloborodov (2006, Fig. 3 there), who got that $\eta_{\rm e}$ always
increases with decreasing $R$ in the very inner region. We think
that $\eta_{\rm e}$ ought to decrease when temperature is very high.
It is important to note that the value of $\eta_{\rm e}$ is of order
a few, i.e., the electron degeneracy is moderate. This justifies
that the exact Fermi-Dirac distribution must be used in the
calculations, and analytic approximations for either extremely
degenerate electrons ($\eta_{\rm e}\gg 1$) or fully nondegenerate
electrons ($\eta_{\rm e} \ll 1$) are invalid for hyperaccretion
disks.

Figure 2 shows the composition of disk matter. In the outer region
between $R = 500R_g$ and $R \sim 200R_g$, almost all
$\alpha$-particles are not disintegrated, so that the
$\alpha$-particles fraction $Y_{\rm \alpha}$  keeps to be $\sim
0.25$, the electron fraction $Y_{\rm e}$ keeps to be $\sim 0.5$, and
the free proton fraction \Yp , the free neutron fraction \Yn , and
the free nucleon fraction $X_{\rm nuc}$  are all keeping $\sim 0$.
From $R \sim 200R_g$ inwards, the disintegration of
$\alpha$-particles causes $Y_{\rm \alpha}$  to decreases and $X_{\rm
nuc}$ to increase dramatically. Because of the neutronization
processes favored by high temperature, $Y_{\rm n}$ greatly exceeds
\Yp , and $Y_{\rm e}$ decreases accordingly. In the innermost region
$R < 10R_g$, $\alpha$-particles are almost fully disintegrated,
i.e., $Y_{\rm \alpha} \sim 0$, $X_{\rm nuc} \sim 1$; and the
neutronization makes $Y_{\rm n}$ larger than 0.9, and $Y_{\rm p}$
and $Y_{\rm e}$ smaller than 0.1.

Contributions to the total pressure $P$ from free nucleons and
$\alpha$-particles $P_{\rm gas}$, from degenerate electrons and
positrons $P_{\rm e}$, from photon radiation $P_{\rm rad}$, and from
neutrinos $P_{\rm \nu}$ are drawn in Figure 3. It is seen that
$P_{\rm e} > P_{\rm gas}$ in the outer region $R \gtrsim 100R_g$,
and $P_{\rm gas}$ becomes dominant at smaller radii because both of
the disintegration of $\alpha$-particles and the decreasing
degeneracy of electrons; $P_{\rm rad}$ is of some importance only in
the outer region, and $P_{\rm \nu}$ is small even in the innermost
region $R < 10R_g$ where the disk becomes optically thick to
neutrinos.

The neutrino optical depth $\tau_{\nu_i}$ is plotted in Figure 4.
Two comparisons are made in Figure 4a. First, the optical depth for
electron neutrinos $\tau_{\nu_{\rm e}}$ is several times larger than
that for $\mu$-neutrinos $\tau_{\nu_{\rm \mu}}$ and $\tau$-neutrinos
$\tau_{\nu_{\rm \tau}}$ , and only $\tau_{\nu_{\rm e}}$ can become
larger than 2/3 in the innermost region $R < 10R_g$, therefore
$\tau_{\nu_{\rm e}}$ ought to be taken as the representative of
$\tau_{\nu_{i}}$. Second, the contribution to $\tau_{\nu_{\rm e}}$
from absorption $\tau_{a,\nu_{\rm e}}$ is more important than that
from scattering $\tau_{s,\nu_{\rm e}}$ , this is a surprising result
as the general understanding in the literature was that scattering
off nucleons is a far more important opacity source than
absorption(e.g., Di Matteo et al. 2002; Lee et al. 2005). We notice
that in calculating the neutrino cooling rate due to the URCA
processes $q_{\rm URCA}$ (which is dominant over other neutrino
cooling rates, see Fig. 4b) those authors used an approximate
formula that is valid only in the nondegeneracy limit. In fact, the
electron degeneracy causes $q_{\rm URCA}$ to increase greatly as in
our calculations. According to equation (46) of Kohri \& Mineshige
(2002), in the complete electron degeneracy limit $q_{\rm URCA}$
becomes extremely large because $q_{\rm URCA} \propto {\eta_{\rm
e}}^9$! It is indeed the case as seen from Figure 4b that the
contribution to $\tau_{a,\nu_{\rm e}}$ from absorption relating to
the URCA processes $\tau_{\rm URCA}$ is most important, the
contribution from neutrino annihilation into $e^{-}e^{+}$ pairs
$\tau_{\rm e^{-}e^{+}}$ is of some relative importance only in the
very outer region, and the contributions from nucleon-nucleon
bremsstrahlung $\tau_{\rm brem}$ and from the inverse process of
plasmon decay $\tau_{\rm plasmon}$ are totally negligible because
the electron degeneracy is not very high. As to $\tau_{s,\nu_{\rm
e}}$ (Fig. 4c), the contribution through scattering off free
neutrons $\tau_{n,\nu_{\rm e}}$ is dominant over other contributions
due to free protons $\tau_{p,\nu_{\rm e}}$ , electrons
$\tau_{e,\nu_{\rm e}}$ , and $\alpha$-particles
$\tau_{\alpha,\nu_{\rm e}}$ ($\tau_{\alpha,\nu_{\rm e}}$ is too
small to be seen in the figure), obviously because of the richness
of neutrons.

Having $\tau_{\nu_{\rm e}}$ in mind, we now check the validity of
our bridging formula for $Y_{\rm e}$, i.e., equation (42). In Figure
5, $Y_{\rm e}$ calculated using equation (42) is drawn by the solid
line. Here $\dot{M} $= 5\Msuns  is taken because only at such a high
accretion rate the inner region of the disk can become noticeably
optically thick. For comparison, the dashed line in the figure draws
$Y_{\rm e}$ in the case that the disk material is assumed to be
globally opaque to neutrinos, which is calculated with (see Eq.
[39]) \beq\ Y_{\rm e}=\frac{1}{2} (1-X_{\rm nuc})+X_{\rm
nuc}{[1+\exp{(\frac{\mu_{\rm e}-Q}{k_{\rm B} T})}]}^{-1}, \eeq and
the dotted line draws $Y_{\rm e}$ if the material is globally
transparent to neutrinos, obtained with the approximate formula of
Beloborodov (2003),\beq\ Y_{\rm e}=\frac{1}{2} (1-X_{\rm
nuc})+X_{\rm nuc} [\frac{1}{2}+0.487 {(\frac{Q/2-\mu_{\rm e}}{k_{\rm
B} T})}]. \eeq These two equations do not include the effect of
varying neutrino optical depth. It is clear that our equation (42)
does represent a bridge between the optically very thick and
optically very thin limits, and it provides a reasonable estimate of
$Y_{\rm e}$ in the intermediate regime where $Y_{\rm e}$  is
overestimated by equation (46) and underestimated by equation (47).

In Figure 6 we show the ratios of various cooling rates to the
viscous heating rate $\Qvis$. The photon radiation cooling is never
important in hyperaccretion disks (it is practically zero when we
try to calculate it). Advective cooling $\Qadv$ is dominant only in
the outer region of the disk $R \gtrsim 200R_g$, because in this
region the photodisintegration of $\alpha$-particles has not started
and the neutrino emission is weak. In the middle region $200R_g
\gtrsim R \gtrsim 50R_g$, the $\alpha$-particle photodisintegration
cooling $\Qphotodis$ dominates, corresponding to a sharp decrease of
$Y_{\rm \alpha}$ and increase of \Xnuc (see Fig. 2). In the inner
region $R \lesssim 50R_g$, the neutrino cooling $\Qnu$ becomes
dominant as expected.

We do not present here numerical results for the disk structure with
varying values of $\alpha$ and $\dot{M}$. Differences caused by
changing these two parameters are only quantitative and have been
discussed in, e.g., Kohri et al. (2005) and Chen \& Beloborodov
(2006). Briefly speaking, $\alpha$ smaller than 0.1 will make the
density higher, the electron degeneracy higher, the electron
fraction lower, and the neutrino-dominated region larger; and at
still larger accretion rates, advection can become dominant over
neutrino cooling again in a very small region ($R < 5R_g$ for
$\dot{M}$=5\Msuns, see Fig. 3  of  Gu et al. 2006), because in that
region the optical depth is very large and neutrinos are trapped in
the disk.

\section{ Neutrino Radiation and Annihilation Luminosities}

Having the neutrino cooling rate $\Qnu$, the neutrino radiation
luminosity (before annihilation) $L_{\rm \nu}$ is obtained as \beq\
L_{\rm \nu}=4 \pi \int_{R_{in}}^{R_{out}} Q_{\rm \nu} R d R. \eeq In
our calculations the inner and outer edge of the disk are taken to
be $R_{in} = 3R_g$ and $R_{out} = 500R_g$, respectively.

For the calculation of the neutrino annihilation luminosity we
follow the approach in Ruffert et al. (1997), Popham et al. (1999),
and Rosswog et al. (2003). The disk is modeled as a grid of cells in
the equatorial plane. A cell $k$ has its mean neutrino energy
$\varepsilon_{\nu_i}^k$, neutrino radiation luminosity $l_{\nu_i}^k$
, and distance to a space point above (or below) the disk $d_k$. The
angle at which neutrinos from cell $k$ encounter antineutrinos from
another cell $k'$ at that point is denoted as $\theta_{kk'}$. Then
the neutrino annihilation luminosity at that point is given by the
summation over all pairs of cells,  \beq\ l_{\nu
\overline{\nu}}=\sum_{i} A_{1,i} \sum_k \frac{l_{\nu_i}^k}{d_k^2}
\sum_{k'} \frac{l_{\overline{\nu}_i}^{k'}}{d_{k'}^2}
(\varepsilon_{\nu_i}^k + \varepsilon_{\overline{\nu}_i}^{k'})
{(1-\cos {\theta_{kk'}})}^2 \nonumber\\+ \sum_{i} A_{2,i} \sum_k
\frac{l_{\nu_i}^k}{d_k^2} \sum_{k'} \frac{l_{\overline
{\nu}_i}^{k'}}{d_{k'}^2} \frac{\varepsilon_{\nu_i}^k +
\varepsilon_{\overline{\nu}_i}^{k'}}{\varepsilon_{\nu_i}^k
\varepsilon_{\overline{\nu}_i}^{k'}} {(1-\cos {\theta_{kk'}})},\eeq
where $A_{1,i} = (1 / 12\pi^2)[\sigma_0/c{(m_{\rm e}
c^2)}^2][{(C_{V,\nu_{i}}-C_{A,\nu_{i}})}^2
+{(C_{V,\nu_{i}}+C_{A,\nu_{i}})}^2]$ , $A_{2,i} = (1/6\pi^2)
(\sigma_0/c)$ $ (2 C_{V,\nu_{i}}^2-C_{A,\nu_{i}}^2)$, with
$C_{V,\nu_{i}}$ and $C_{A,\nu_{i}}$ given in \S 2.3.1. The total
neutrino annihilation luminosity is obtained by the integration over
the whole space outside the black hole and the disk, \beq\ L_{\nu
\overline{\nu}}=4 \pi \int_{R_g}^\infty \int_H^\infty l_{\nu
\overline{\nu}} R d R d Z. \eeq

Figure 7 shows $L_{\rm \nu}$ (the thick dashed line) and  $L_{\nu
\overline{\nu}}$ (the thick solid line) with varying $\dot{M}$ ($M =
3 M_\sun$, $\alpha = 0.1$, and $j = 1.8 c R_g$ are kept). For
comparison, these two luminosities calculated in our previous work
(Gu et al. 2006), where the electron degeneracy was not correctly
considered and $Y_{\rm e}$  was taken to be equal to 0.5, are also
given in the figure by the thin dashed line and thin solid line,
respectively. It is clear that the electron degeneracy and the lower
$Y_{\rm e}$ resulted from the neutronization processes indeed
suppress the neutrino emission considerably, the resulting $L_{\rm
\nu}$ and $L_{\nu \overline{\nu}}$  are reduced by a factor $\sim
30\% - 70\%$ comparing with their overestimated values in Gu et al.
(2006). Even so, the correct $L_{\nu \overline{\nu}}$  is still well
above $10^{50}~{\rm ergs~s^{-1}}$ provided $\dot{M} \sim$ 1\Msuns,
and reaches to $\sim 10^{52}~{\rm ergs~s^{-1}}$ when  $\dot{M}\sim$
10\Msuns. Therefore, based on the energy consideration,
neutrino-cooled accretion disks can work as the central engine of
GRBs. Note that our calculations are for a nonrotating black hole,
both Popham et al. (1999) and Chen \& Beloborodov (2006) have shown
that a spinning (Kerr) black hole will enhance the neutrino
radiation efficiency, this only strengthens our conclusion here.

In Figure 8 the corresponding neutrino radiation efficiency
$\eta_{\rm \nu} (\equiv L_{\rm \nu}/\dot{M }c^2)$ and neutrino
annihilation efficiency $\eta_{\nu \overline{\nu}} (\equiv {L_{\nu
\overline{\nu}}}/{L_{\rm \nu}}$ ) are shown by the dashed line and
solid line, respectively. It is seen that $\eta_{\rm \nu}$ does not
change much with varying $\dot{M}$; while $\eta_{\nu
\overline{\nu}}$ increases rapidly with increasing $\dot{M}$, this
is because at higher accretion rates more neutrinos are emitted and
in turn have higher probabilities to encounter with each other.

To see the spatial distribution of neutrino annihilation luminosity,
we plot in Figure 9 contours of ($2\pi R  l_{\nu \overline{\nu}}$)
in units of (${\rm erg~s^{-1}~cm^{-2}}$), i.e., the neutrino
annihilation luminosity of a circle with cylindrical coordinates $R$
and $Z$, for $\dot{M}$= 1\Msuns. The figure has two important
implications. First, it demonstrates the strong focusing of neutrino
annihilation towards the central region of space. In fact, by
performing the integration of equation (50) out to each radius we
get that nearly 60\% of the total annihilation luminosity is ejected
from the region $R < 20R_g$. Second, it shows that the annihilation
luminosity varies more rapidly along the $Z$ coordinate than along
the $R$ coordinate, indicating that this luminosity is anisotropic
and most of the annihilation energy escapes outward along the
angular momentum axis of the disk.

\section{Summary and Discussion}

In this paper we wish to discuss whether the annihilation of
neutrinos emitted from hyperaccretion disks can provide sufficient
energy for GRBs, i.e., to estimate the neutrino annihilation
luminosity $L_{\nu \overline{\nu}}$. To do this, we need to know the
neutrino optical depth $\tau_{\nu_i}$, because it determines the
neutrino cooling rate $\Qnu$ (eq. [9]) and the neutrino radiation
luminosity $L_{\nu}$ (eq. [48]).

To calculate contributions  to $\tau_{\nu_i}$ from various
absorption and scattering processes of neutrinos (eq. [16]), we need
to know the composition and physical state of disk matter, namely
the fractions of electrons $Y_{\rm e}$  and free nucleons \Xnuc ,
and the electron chemical potential $\mu_{\rm e}$ (or electron
degeneracy $\eta_{\rm e}$), with given density $\rho$ and
temperature $T$. We give exact definitions of the free proton
fraction $Y_{\rm p}$  and free neutron fraction $Y_{\rm n}$  and
their relations to $Y_{\rm e}$ and $X_{\rm nuc}$, and get three
equations (eqs. [42], [44], and [45]) from the conditions of
$\beta$-equilibrium, nuclear statistical equilibrium, and charge
neutrality that describe the interdependence of $Y_{\rm p}$, $X_{\rm
nuc}$, $\mu_{\rm e}$, and $\tau_{\nu_i}$ .

We prove that the electron degeneracy has important effects indeed,
but mainly not in the sense that it suppresses the creation of
neutrinos from ${\rm e^{-}e^{+}}$ pairs and enlarges the electron
pressure (Kohri \& Mineshige 2002; Kohri et al. 2005). As seen from
Figures 4b and 3, the neutrino cooling due to ${\rm e^{-}e^{+}}$
pair annihilation $q_{\rm e^{-}e^{+}}$ is much smaller than that due
to the URCA processes $q_{\rm URCA}$, and the electron pressure is
important only in the outer region of the disk where the
$\alpha$-particle disintegration has not started. Instead, the main
effects of electron degeneracy are: (1) It increases $q_{\rm URCA}$
greatly, so that the corresponding absorption makes a major
contribution to the neutrino optical depth (Fig. 4); (2) It along
with the neutronization processes cause $Y_{\rm e}$  to become
smaller than 0.1 in the inner region of the disk (Fig. 2) where the
neutrino cooling is dominant (Fig. 6).

The resulting $L_{\nu \overline{\nu}}$ is considerably reduced
comparing with that in previous works where the electron degeneracy
was not considered and $Y_{\rm e}$  was taken to be 0.5, however it
is still likely to be adequate for GRBs, and its spatial
distribution is likely to be anisotropic.

Perhaps the main limitation of our calculations here is that they
are one-dimensional. As in most of previous works in the field, we
do not study the vertical structure of the disk and instead use the
simple 'one-zone' approximation of the vertically-averaged model. In
particular, we do not consider the distribution and transport of
neutrinos in the vertical direction of the disk, a problem that
remains unsolved or even rarely touched (Sawyer 2003). Strictly
speaking, a reliable quantitative evaluation of the neutrino
annihilation luminosity should require two-dimensional calculations,
in which the vertical structure and neutrino transport are
self-consistently included. But in view of the fact that most
popular models of normal accretion disks (the Shakura-Sunyaev disk
model, the slim disk model, and the advection-dominated accretion
flow model) are also one-dimensional and have been proved
successful, our results here may provide a plausible, though rough,
estimate of the luminosity of neutrino-cooled accretion disks.

\acknowledgments

This work was supported by the National Science Foundation of China
under grants 10503003 and 10673009.

\clearpage
\begin{figure}
\plotone{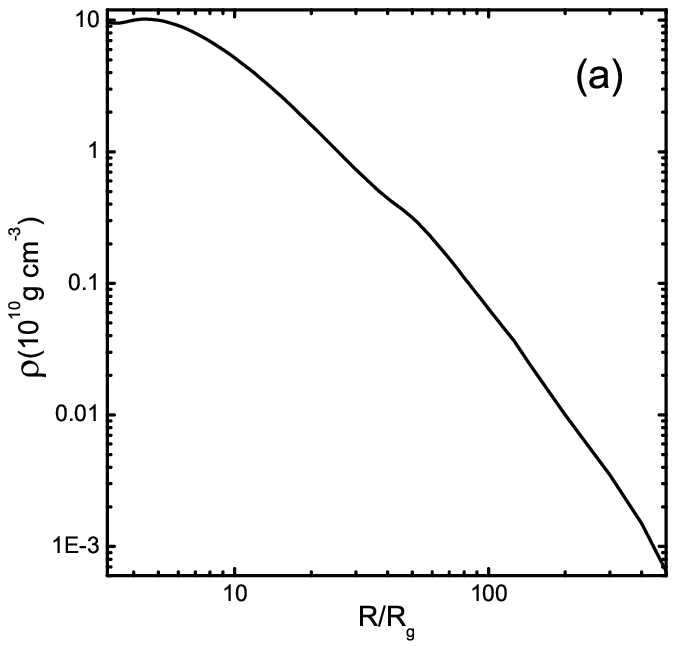}
\end{figure}

\begin{figure}
\plotone{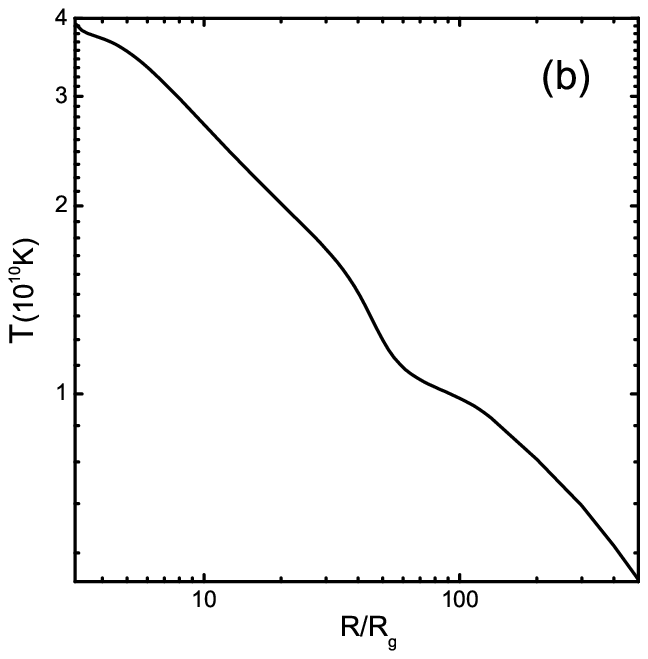}
\end{figure}

\begin{figure}
\plotone{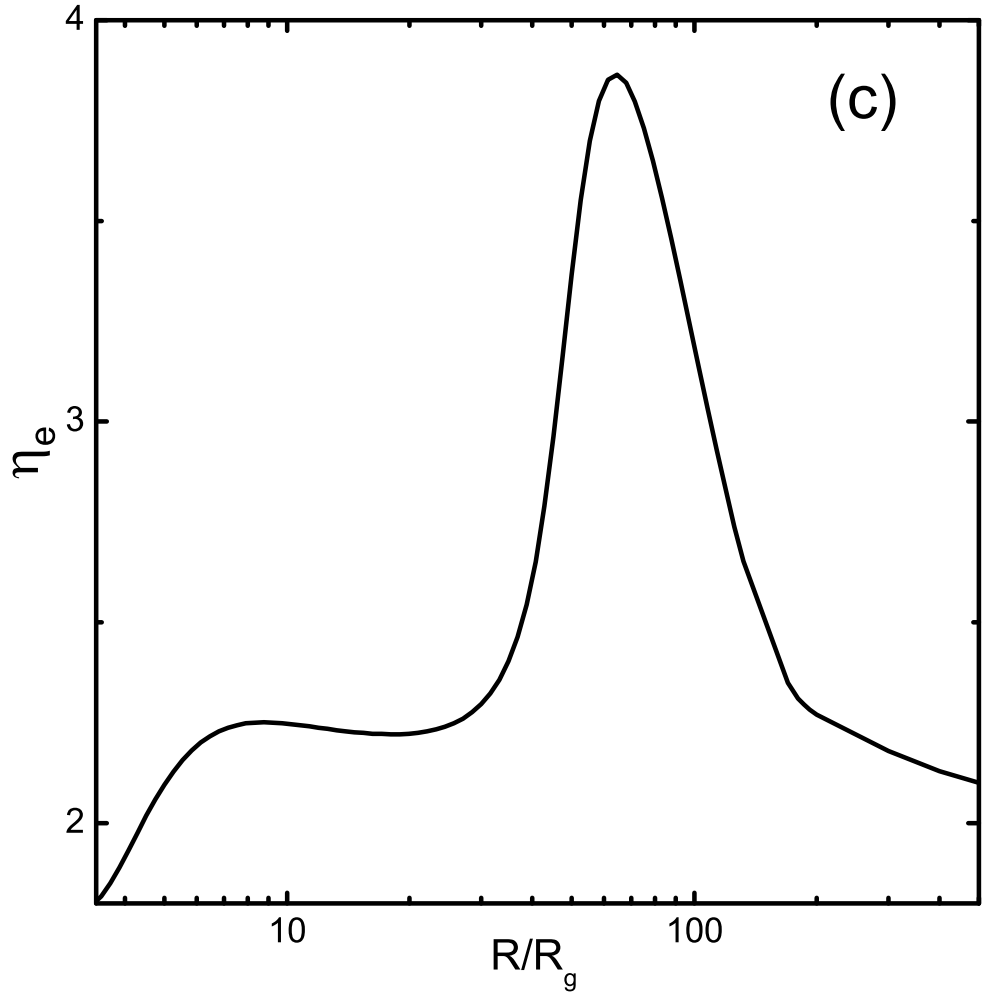} \caption{ (a) Density $\rho$, (b) Temperature $T$,
and (c) electron degeneracy $\eta_{\rm e}$ as functions of radius
$R$, with the black hole mass $M = 3M_\sun$, mass accretion rate
$\dot{M}$ = 1\Msuns, viscosity parameter $\alpha = 0.1$, and the
accreted specific angular momentum $j = 1.8 c R_g$. \label{fig1}}
\end{figure}
\clearpage

\begin{figure}
\plotone{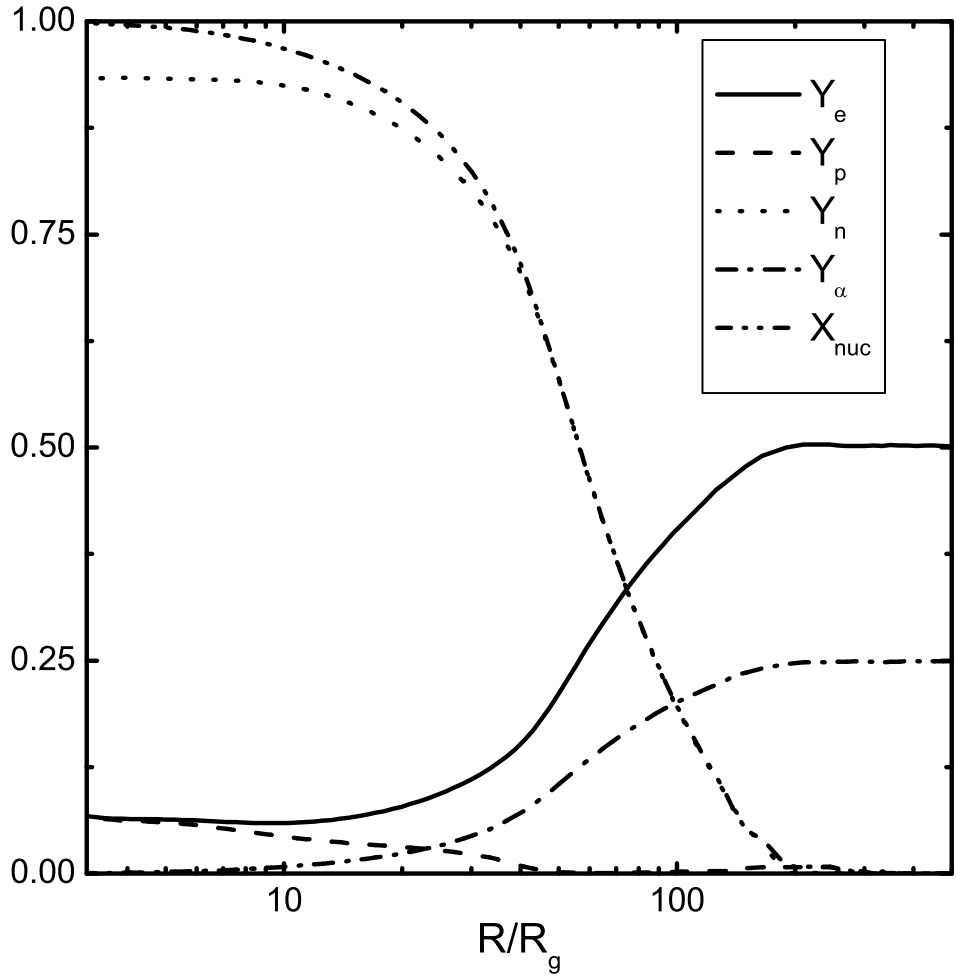} \caption{ Electron fraction \Ye , free proton
fraction \Yp , free neutron fraction \Yn , $\alpha$-particle
fraction \Yalpha , and free nucleon fraction $X_{\rm nuc}$  as
functions of $R$, with the same constant parameters as in Fig. 1.
\label{fig2}}
\end{figure}

\clearpage

\begin{figure}
\plotone{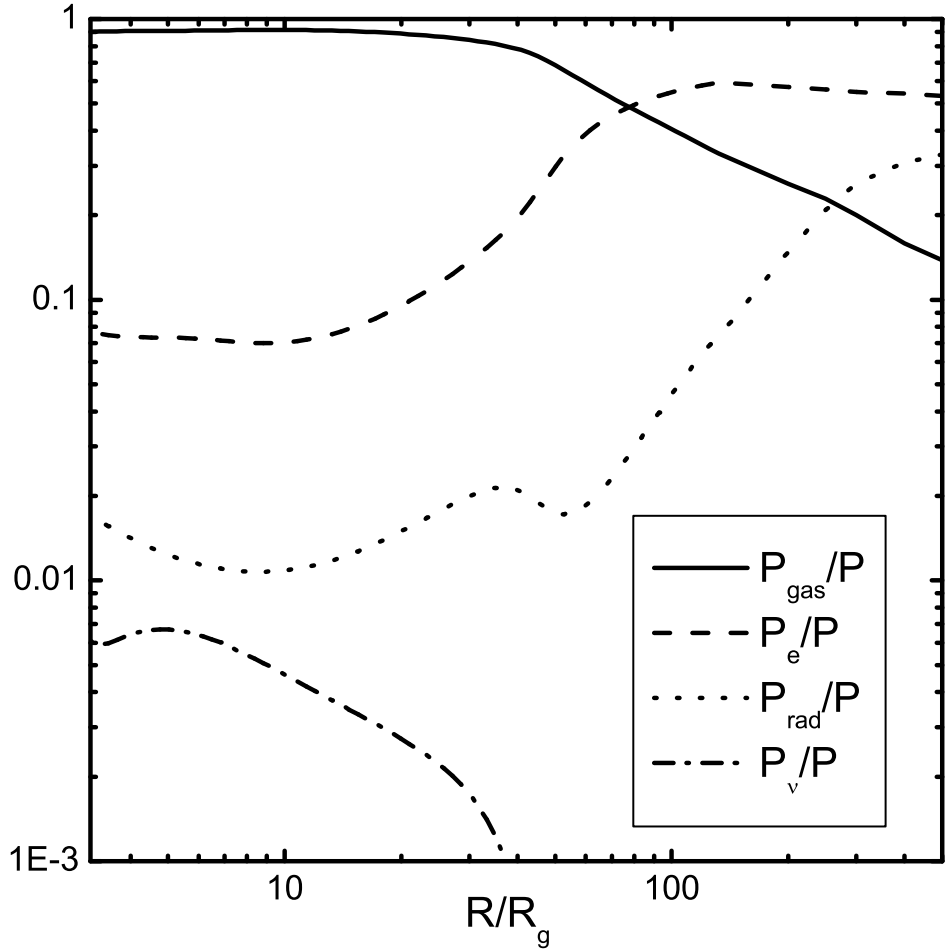} \caption{ Contributions to the total pressure $P$
from free nucleons and $\alpha$-particles $P_{\rm gas}$, from
degenerate electrons $P_{\rm e}$, from photon radiation $P_{\rm
rad}$, and from neutrinos $P_{\rm \nu}$ as functions of $R$, with
the same constant parameters as in Fig. 1. \label{fig3}}
\end{figure}

\clearpage
\begin{figure}
\plotone{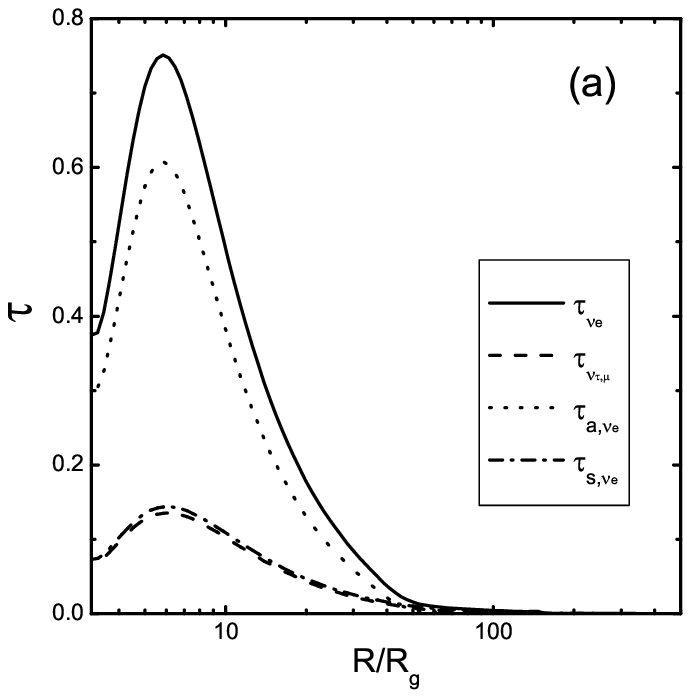}
\end{figure}

\begin{figure}
\plotone{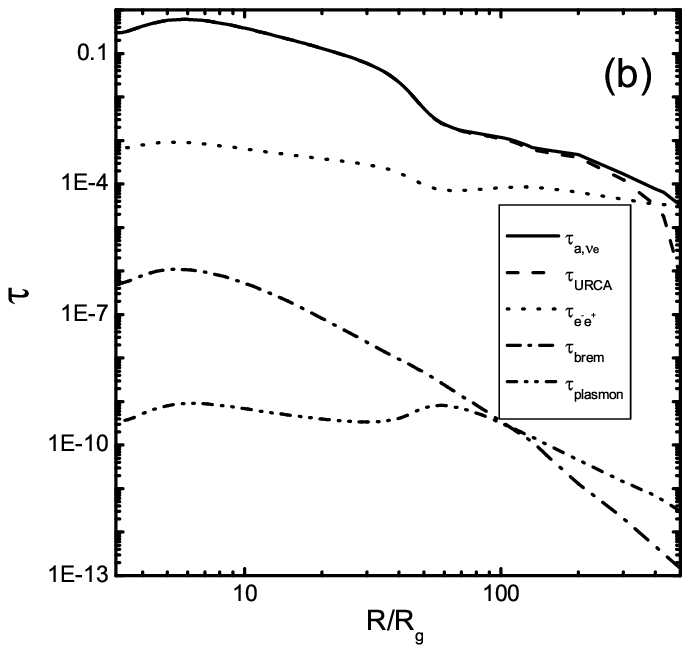}
\end{figure}

\begin{figure}
\plotone{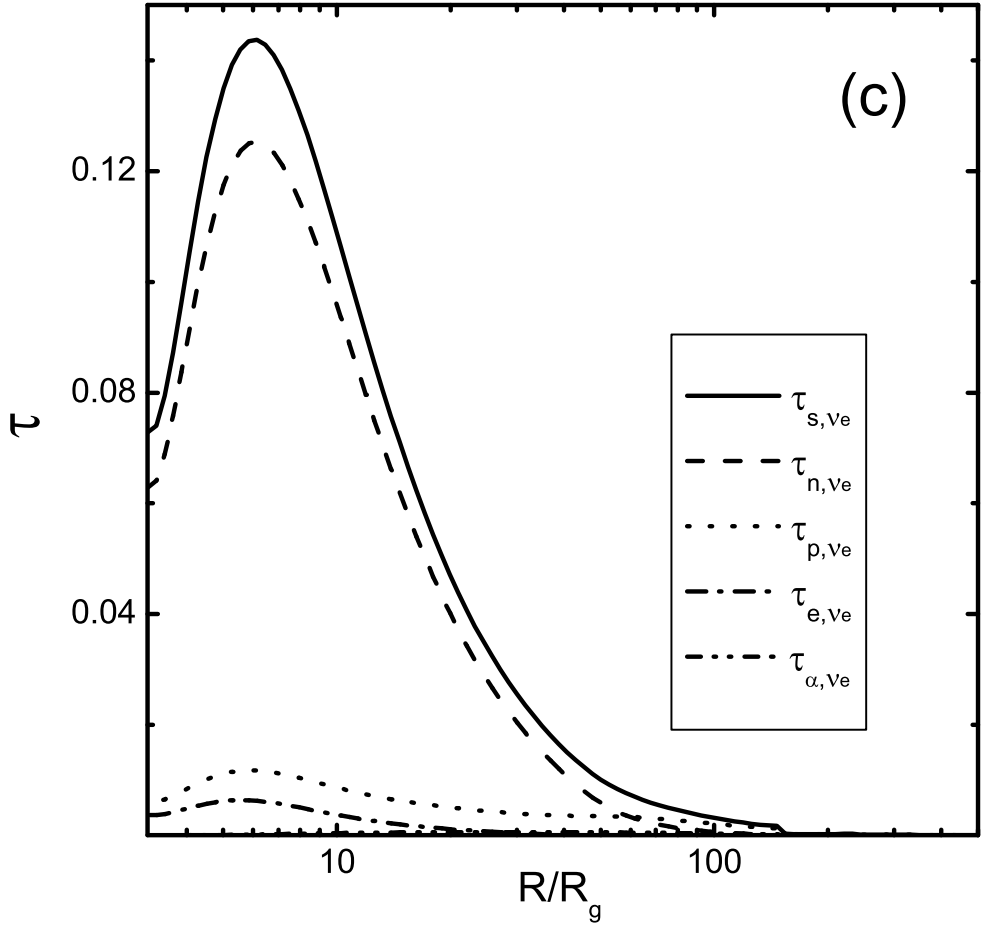} \caption{ (a) Total optical depth for electron
neutrinos $\tau_{\nu_e}$ , total optical depth for $\tau$-neutrinos
and $\mu$-neutrinos $\tau_{\nu_{\tau,\mu}}$ , absorption optical
depth for electron neutrinos $\tau_{a,\nu_{\rm e}}$ , and scattering
optical depth for electron neutrinos $\tau_{s,\nu_{\rm e}}$ ; (b)
Quantity $\tau_{a,\nu_{\rm e}}$ and its contributions from the URCA
processes $\tau_{\rm URCA}$, from neutrino annihilation $\tau_{\rm
e^{-}e^{+}}$ , from nucleon-nucleon bremsstrahlung $\tau_{\rm
brem}$, and from the inverse process of plasmon decay $\tau_{\rm
plasmon}$; and (c) Quantity $\tau_{s,\nu_{\rm e}}$ and its
contributions due to free neutrons $\tau_{n,\nu_{\rm e}}$ , due to
free protons $\tau_{p,\nu_{\rm e}}$ , due to electrons
$\tau_{e,\nu_{\rm e}}$ , and due to $\alpha$-particles
$\tau_{\alpha,\nu_{\rm e}}$ as functions of $R$, with the same
constant parameters as in Fig. 1. \label{fig4}}
\end{figure}

\begin{figure}
\plotone{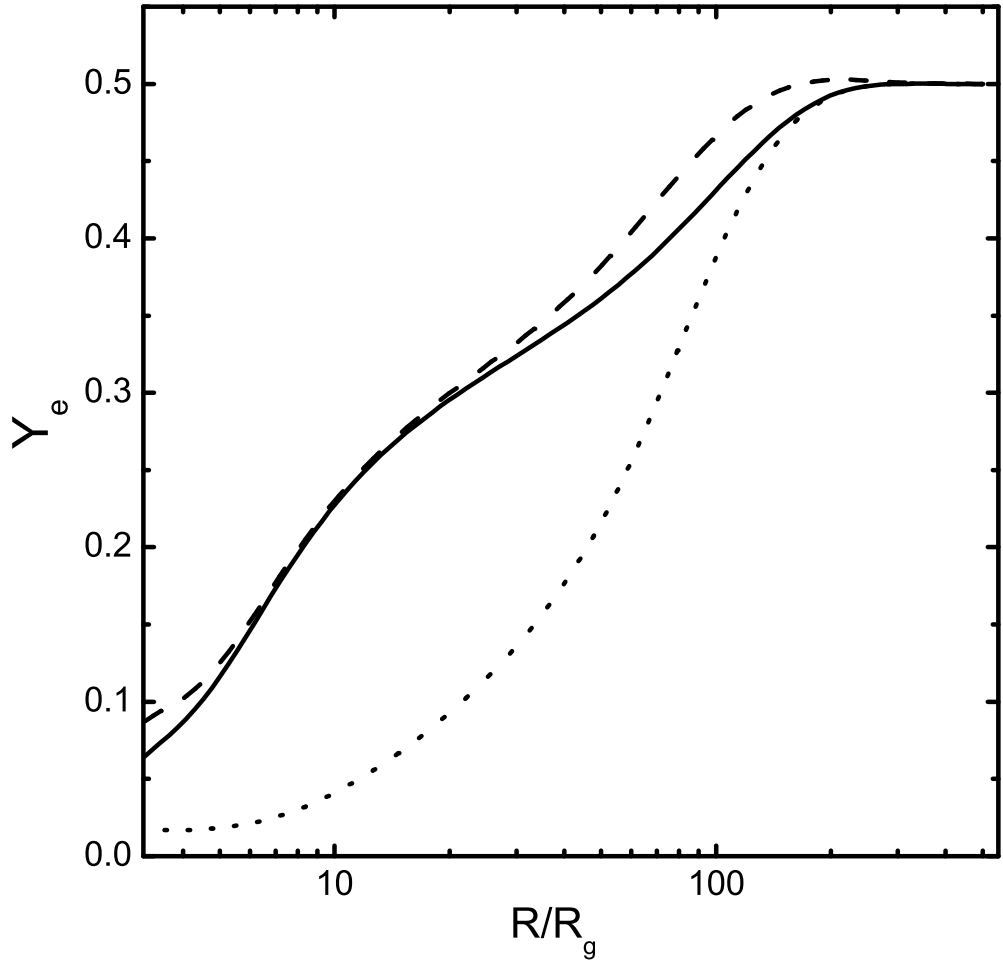} \caption{ Solid, dashed, and dotted lines draw
$Y_{\rm e}$ calculated with equations (42), (46), and (47),
respectively, with the same constant parameters as in Fig. 1 except
for$\dot{M} = 5$\Msuns. \label{fig5}}
\end{figure}

\begin{figure}
\plotone{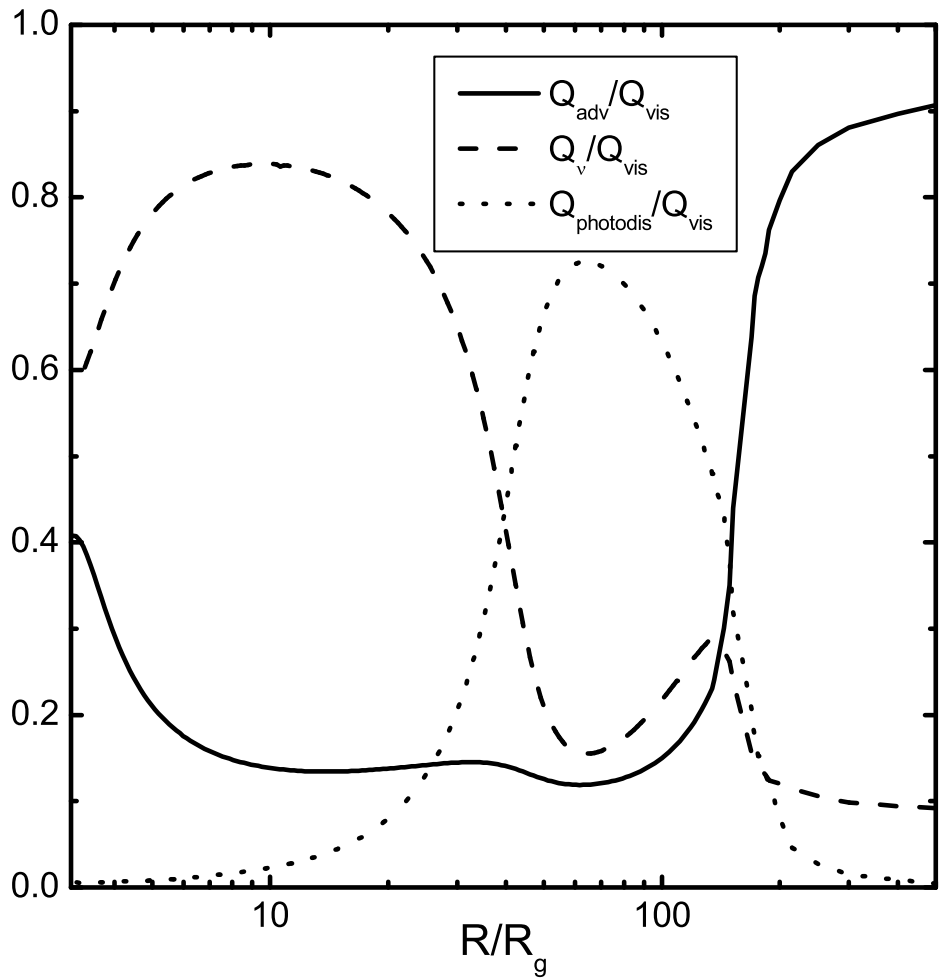} \caption{ Ratios of the advective cooling rate
$\Qadv$, $\alpha$-particle photodisintegration cooling rate
$\Qphotodis$ , and neutrino cooling rate $\Qnu$ to viscous heating
rate $\Qvis$ as functions of $R$, with the same constant parameters
as in Fig. 1. \label{fig6}}
\end{figure}

\begin{figure}
\plotone{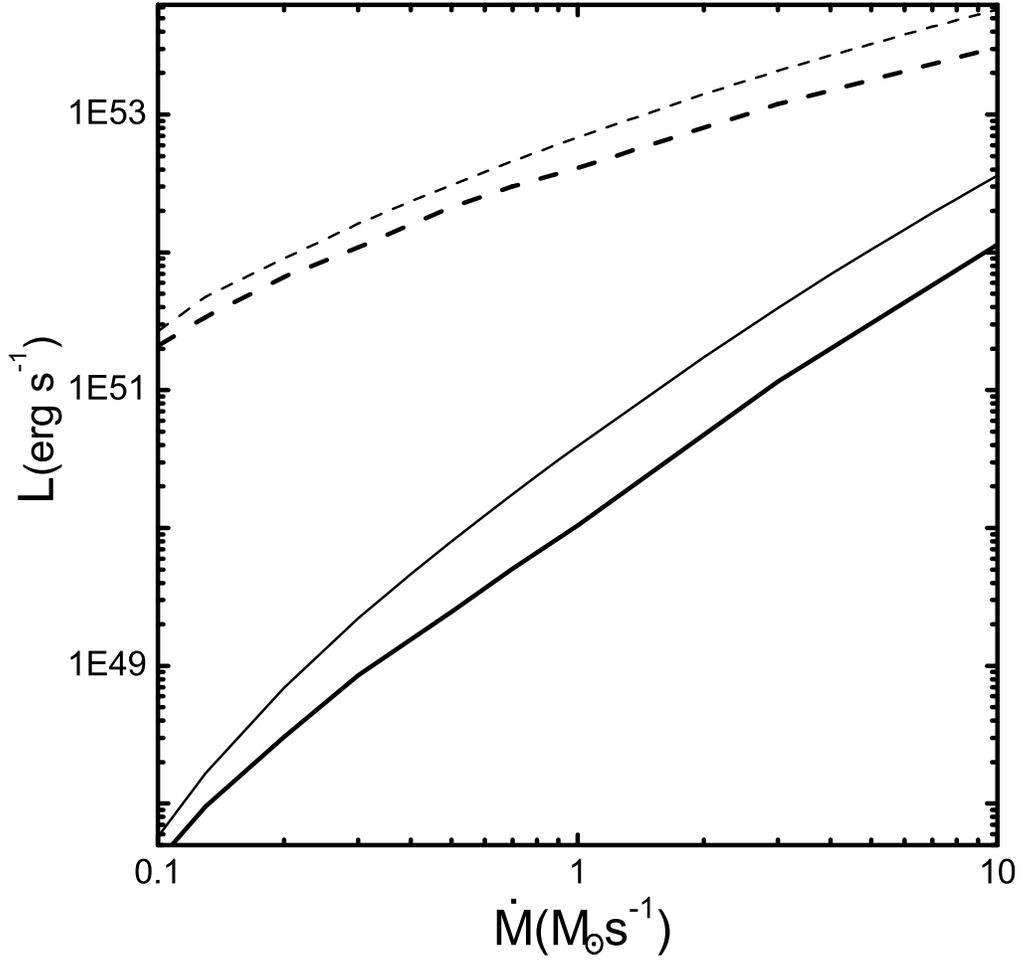} \caption{ Neutrino radiation luminosity $L_{\rm
\nu}$ (the thick dashed line) and neutrino annihilation luminosity
$L_{\nu \overline{\nu}}$ (the thick solid line) for varying
$\dot{M}$. The overestimated $L_{\rm \nu}$ (the thin dashed line)
and $L_{\nu \overline{\nu}}$ (the thin solid line) taken from Gu et
al. (2006) are also given. The constant parameters $M$, $\alpha$,
and $j$ are the same as in Fig. 1. \label{fig7}}
\end{figure}

\begin{figure}
\plotone{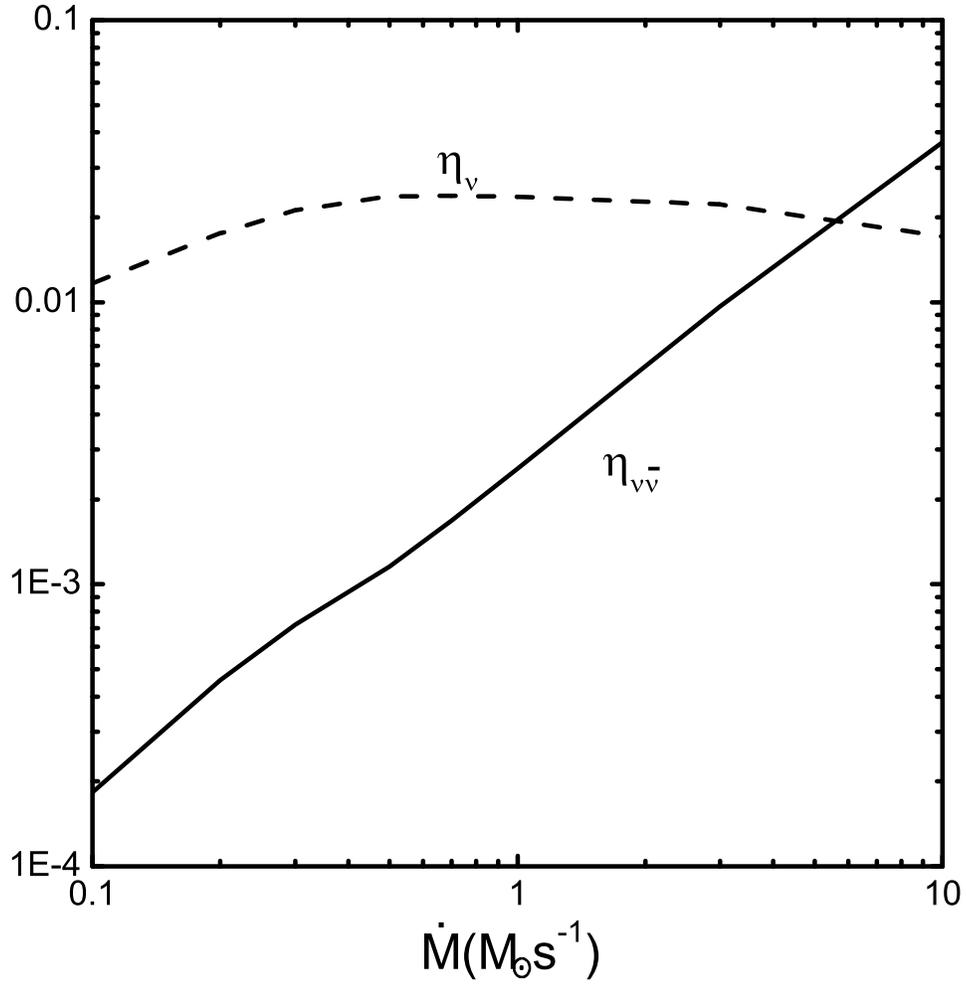} \caption{ Neutrino radiation efficiency $\eta_{\rm
\nu}$ and neutrino annihilation efficiency $\eta_{\nu
\overline{\nu}}$ for varying $\dot{M}$ , with the same $M$,
$\alpha$, and $j$ as in Fig. 1. \label{fig8}}
\end{figure}

\begin{figure}
\plotone{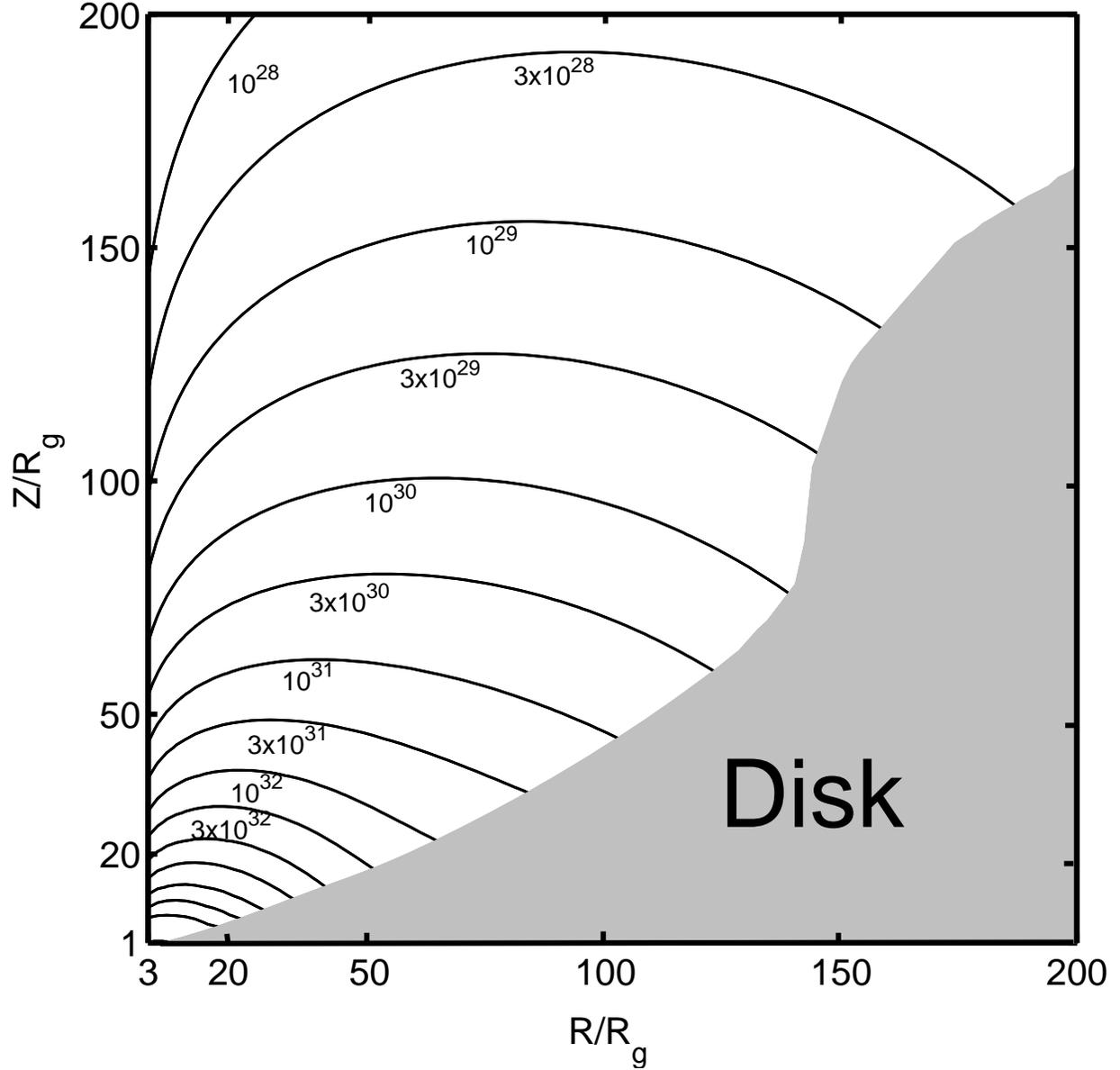} \caption{ Contours of the neutrino annihilation
luminosity of a circle with cylindrical coordinates $R$ and $Z$. The
number attaching to each line is this luminosity in units of (${\rm
erg~s^{-1}~cm^{-2}}$). The shaded region shows the accretion disk.
The constant parameters are the same as in Fig. 1.  \label{fig9}}
\end{figure}

\clearpage

\end{document}